# Using large language models to promote health equity

Emma Pierson[1,2,14], Divya Shanmugam[3,14], Rajiv Movva[1,14], Jon Kleinberg[4,14], Monica Agrawal[5], Mark Dredze[6], Kadija Ferryman[6], Judy Wawira Gichoya[7], Dan Jurafsky[8], Pang Wei Koh[9], Karen Levy[4], Sendhil Mullainathan[10], Ziad Obermeyer[11], Harini Suresh[12], Keyon Vafa[13]

1: Cornell Tech, New York, NY, USA. 2: Weill Cornell Medical College, New York, NY, USA. 3: Massachusetts Institute of Technology, Cambridge, MA, USA. 4: Cornell University, Ithaca, NY, USA. 5: Duke University, Durham, NC, USA. 6: Johns Hopkins University, Baltimore, MD, USA. 7: Emory University, Atlanta, GA, USA. 8: Stanford University, Palo Alto, CA, USA. 9: University of Washington, Seattle, WA, USA. 10: University of Chicago, Chicago, IL, USA. 11: University of California, Berkeley, Berkeley, CA, USA. 12: Brown University, Providence, RI, USA. 13: Harvard University, Cambridge, MA, USA. 14: Denotes co-first authorship. Correspondence to: emmapierson@berkeley.edu.

**Abstract**

While the discussion about the effects of large language models (LLMs) on health equity has been largely cautionary, LLMs also present significant opportunities for improving health equity. We highlight three such opportunities: improving the detection of human bias; creating structured datasets relevant to health equity; and improving equity of access to health information.

**Description**

Advances in large language models (LLMs) have expanded the possibilities for addressing health equity, presenting opportunities to complement ongoing efforts to mitigate biases. This paper explores three promising applications of LLMs — bias detection, structured data creation, and equitable health information access — while emphasizing the importance of centering underserved populations in their development and deployment.

Interest in the societal impacts of large language models (LLMs) has increased due to recent improvements in their capabilities. Discussions about how they will affect health equity have been largely cautionary (Figure 1), focusing on questions such as, "How might LLMs be biased, and how would we mitigate those biases?" This is a vital conversation; The ways that AI — and LLMs specifically — can entrench biases have been extensively documented.[1,2] But equally vital, and much less discussed, is the more opportunity-focused counterpoint: "What promising applications do LLMs enable that could *promote* health equity?"

A long tradition of efforts to promote equitable AI, both in health and more broadly, shows it is essential for researchers and policymakers to take this broader view, seeking out equity-promoting use cases from the outset.[3] There are many choices that determine the impacts of AI, and a fundamental choice very early in the pipeline is the *problems we choose to apply it to*. If we focus only later in the pipeline — making LLMs somewhat more fair as they facilitate primary use cases that intrinsically entrench power — we will miss an important opportunity to guide them to equitable impacts. Past research has highlighted the many equity-promoting use cases of AI[4] such as formalizing social problems[5], supporting the work of activists[6], and diagnosing biases.[7] Past health equity research has also shown that, without a deliberate focus on equity-promoting use cases, the needs of underserved populations will be underfunded and overlooked.[3] We build on this tradition of work to explore such possibilities for LLMs.

Rapid recent improvements in LLMs mean that the set of possible health-equity-promoting use cases involving language has expanded dramatically and has yet to be fully explored. We highlight the emerging potential of LLMs to promote health equity by presenting three newly possible, promising research directions while keeping risks in clear view. We draw on extensive conversations with AI researchers, healthcare practitioners, and policymakers who see potential in using LLMs to promote health equity.

## Equity-promoting applications of LLMs

**Application 1: LLMs can improve the detection of bias**

Human biases contribute to inequity. For example, doctors describe Black patients as "difficult" more often than white patients[8], a textual indicator of the stigmatization and bias which can pervade the medical pipeline and contribute to health disparities. LLMs can improve our ability to detect human bias from text data. For example, previous research finding evidence of biased language in clinical notes has often relied on relatively rudimentary linguistic analysis methods, like searching for specific words. Relative to prior methods, LLMs can more precisely characterize subtle linguistic biases (e.g., in sentiment, tone, and stereotypes) that rely on understanding the surrounding context. Recent work has highlighted the value of using LLMs to identify such biases within medicine (e.g., racialized doubts of patient credibility and compliance).[9] (Importantly, LLMs themselves have biases[1], but past work has shown that they can nonetheless serve as useful diagnostics for bias.[9]) Similar LLM-based techniques could accelerate the detection of biases in medical textbooks, which often involves time-consuming manual annotation by domain experts. All these applications flow from the fact that rich text datasets, like clinical notes and medical textbooks, frequently document human behavior. LLMs can characterize biases in behavior detectable from text. The approach treats text data as an "artifact"[7] created by a biased health ecosystem, which can be examined to understand and mitigate biased practices. This echoes calls for researchers to "study up" — focusing their analysis on the powerful actors, structures, and institutions perpetuating disparities.[5]

**Application 2: LLMs can create structured datasets relevant to health equity**
Structured datasets, which track a standardized set of fields (e.g., the race and health insurance provider of a patient), allow us to quantify health disparities and develop policies to mitigate them. LLMs, which have shown great promise in structured information extraction[10], can reduce the human effort required to create these structured datasets from raw text data. For example, healthcare researchers manually extract employment, housing, and insurance measures from clinical notes to understand the impact of these factors on patient outcomes. Clinicians manually extract details on the demographics of clinical trial participants from unstructured trial reports to assess trial diversity. Activists collect news articles to create databases tracking public health threats that disproportionately impact marginalized populations, like gender-based violence. Journalists collate legislative texts to produce structured databases of restrictions on women's and reproductive healthcare.

These use cases have traditionally required either extensive manual effort to extract structured data by hand or custom-built engineering pipelines. In contrast, LLMs can extract structured information from unstructured datasets, even with limited training and very little technical modification, in domains where they were not specifically trained[10], dramatically broadening both the settings in which these strategies can be applied and the participants who can lead these efforts. As with all applications we discuss here, researchers need to center the communities they seek to support. For example, LLM tools that track gender-based violence should be co-designed with activists across many countries who have extensive involvement in this work in order to ensure their true needs are being addressed.[6,11]

**Application 3: LLMs can improve access to health information**
Significant inequities persist in access to specialized health information. For patients who lack easy access to a medical provider or cannot afford to see one, LLMs can be used to help provide accurate and understandable answers to patient questions[12], empowering patients to advocate for their health needs. In contrast to traditional internet search, which can only direct users to pre-existing webpages, LLMs can produce answers tailored to patient questions from multiple sources, conversationally and, when relevant, cross-lingually. Indeed, there are ongoing global health efforts to build and deploy LLM-based systems in developing countries to alleviate gaps in patient access to health information.[13] For patients and caregivers who do interact with the healthcare system, LLMs can be applied to translate clinical jargon[13] within diagnostic reports or clinical notes into easy-to-understand language. LLMs are still not a replacement for clinicians[14] and should not be used as a magic bullet displacing existing efforts to improve access to care[15], but rather as a supplement on top of these efforts which expands patient access to health-related information.

## Steps towards implementation

As the research community's discussion of the impact of AI on equity has centered on risks, the policy landscape has largely followed suit; AI research has been framed as having significant potential for growth but potentially harmful effects on marginalized groups. As such, policy has focused on audits and regulation with an eye toward possible harms. Our main policy recommendation thus echoes our call for researchers: policymakers should seek not only to reduce equity-related harms, but also to incentivize equity-promoting use cases. Markets are unlikely to provide adequate rewards for tools that benefit underserved populations; Policy levers like research funding and procurement can be used to provide adequate incentives.

For example, consider one of our proposed use cases: building a medical conversational AI system to broaden access to health information for underserved populations. Policymakers can incentivize the development of such equity-promoting systems via research funding and procurement calls. Research funding could incentivize evaluations that center equity: conversational AI systems need to be tested on actual queries from diverse patient populations, including underserved patient populations; perform well in multiple languages; and avoid caricaturing or stereotyping patients.[2] This should happen in parallel with basic research funding to mitigate errors like hallucinations and to fine-tune general LLMs to create health-specific models. Finally, because health inequity stems from myriad socioeconomic, environmental, and structural factors[16] that defy simple technical fixes, LLMs are only one tool in the toolbox, not a panacea for health inequity.

However, the use cases we highlight suggest that LLMs are potent tools which can improve bias detection, create structured datasets relevant to health equity, and improve access to health information. We stand to gain enormously if we recognize these applications by intervening in the early stages of design and advancing the interests of underserved populations.

# References

1. Weidinger L, Uesato J, Rauh M, et al. Taxonomy of risks posed by language models. Proceedings of the 2022 ACM Conference on Fairness, Accountability, and Transparency. 2022:214-229.

2. Zack T, Lehman E, Suzgun M, et al. Assessing the potential of GPT-4 to perpetuate racial and gender biases in health care: a model evaluation study. Lancet Digit Health. 2024;6(1)

3. Chen IY, Pierson E, Rose S, Joshi S, Ferryman K, Ghassemi M. Ethical machine learning in healthcare. Annu Rev Biomed Data Sci. 2021;4(1):123-144.

4. Rodriguez JA, Alsentzer E, Bates DW. Leveraging large language models to foster equity in healthcare. J Am Med Inform Assoc. 2024; 31(9): 2147–2150.

5. Abebe R, Barocas S, Kleinberg J, Levy K, Raghavan M, Robinson DG. Roles for computing in social change. Proceedings of the 2020 Conference on Fairness, Accountability, and Transparency. 2020:252-260.

6. Suresh H, Movva R, Dogan AL, et al. Towards intersectional feminist and participatory ML: a case study in supporting feminicide counterdata collection. Proceedings of the 2022 ACM Conference on Fairness, Accountability, and Transparency. 2022:667-678.

7. Ferryman K, Mackintosh M, Ghassemi M. Considering biased data as informative artifacts in AI-assisted health care. N Engl J Med. 2023;389(9):833-838.

8. Sun M, Oliwa T, Peek ME, Tung EL. Negative patient descriptors: documenting racial bias in the electronic health record. *Health Aff (Millwood)*. 2022;41(2):203-211. https://www.healthaffairs.org/doi/10.1377/hlthaff.2021.01423. 10.1377/hlthaff.2021.01423

9. Harrigian K, Zirikly A, Chee B, Ahmad A, Links A, Saha S, Beach MC, Dredze M. Characterization of stigmatizing language in medical records. *Proceedings of the 61st Annual Meeting of the Association for Computational Linguistics (Volume 2: Short Papers)*. 2023:312-329.

10. Agrawal M, Hegselmann S, Lang H, Kim Y, Sontag D. Large language models are few-shot clinical information extractors. *Proceedings of the 2022 Conference on Empirical Methods in Natural Language Processing*. 2022:1998-2022.

11. Dulka A. The use of artificial intelligence in international human rights law. Stanford Technology Law Review. 2022;26:316.

12. Yalamanchili A, Sengupta B, Song J, et al. Quality of large language model responses to radiation oncology patient care questions. *JAMA Netw Open*. 2024;7(4)

13. Lyu Q, Tan J, Zapadka ME, et al. Translating radiology reports into plain language using ChatGPT and GPT-4 with prompt learning: results, limitations, and potential. *Vis Comput Ind Biomed Art*. 2023;6(1):9.

14. Hager P, Jungmann F, Holland R, et al. Evaluation and mitigation of the limitations of large language models in clinical decision-making. *Nat Med*. 2024;30(9):2613-2622.

15. Shaffer J, Alenichev A, Faure MC. The Gates Foundation's new AI initiative: attempting to leapfrog global health inequalities? *BMJ Glob Health*. 2023;8(11)

16. Baciu A, Negussie Y, Geller A, Weinstein JN, eds. Communities in action: pathways to health equity. 2017.

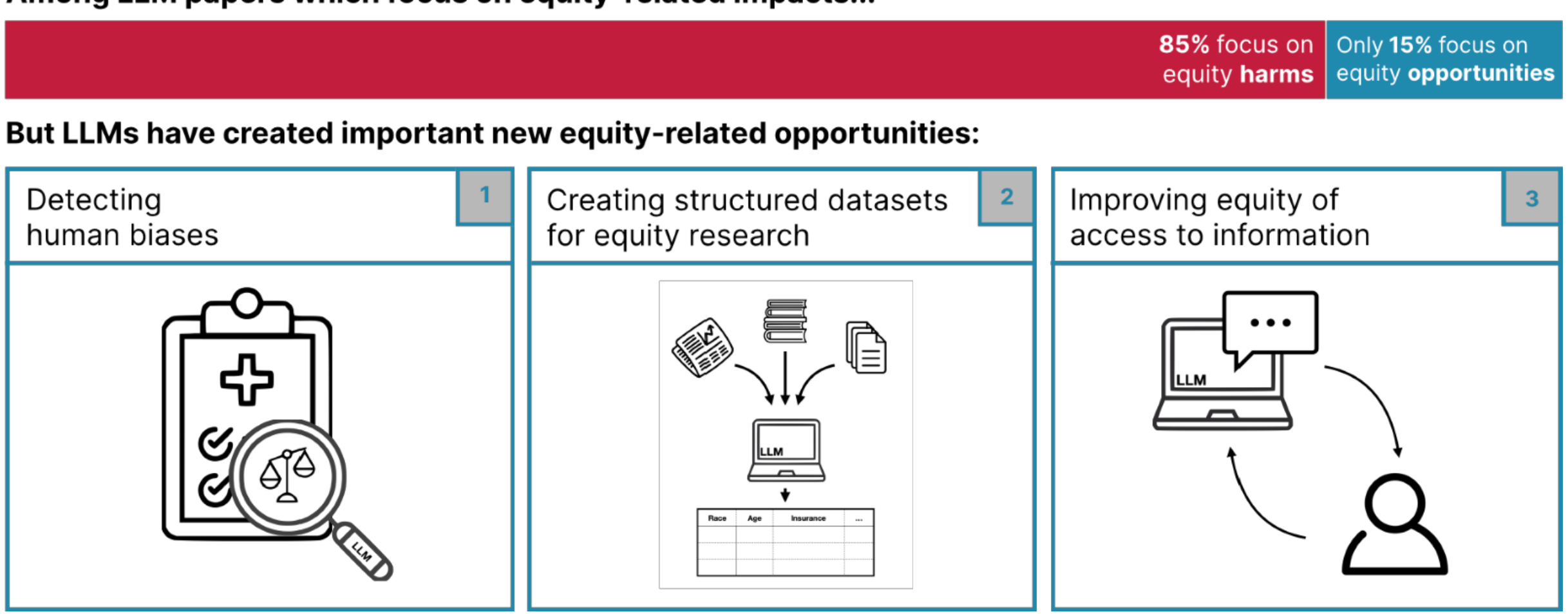

# Figure Captions

Figure 1: Many more LLM papers focus on equity-related harms than opportunities (see SI for methods). We discuss three significant opportunities for using LLMs to promote health equity.

# Supplementary Information

We quantified the extent to which equity-related LLM research papers focused on equity-promoting opportunities vs. mitigating equity-related harms. Using the Semantic Scholar Open Research Corpus (*1*), we identified 27,019 papers written between January 2018 and October 2023 that mentioned a language modeling-related keyword in their title or abstract, following the methodology described in (*2*).[1] Out of this sample, 2,016 papers also mentioned at least one of the following equity-related keywords: {'fairness', 'bias', 'equity', 'equitable', 'ethical', 'justice', 'disparities'}. Out of these 2,016 papers, 100 papers were randomly sampled and independently annotated by three study authors for two characteristics based on their title and abstract:

1. Whether the paper was in fact equity-related (to avoid the problem that some papers used an equity keyword in a non-equity-related context, e.g. "inductive bias.") A paper was classified as equity-related only if equity impacts were a substantial focus of the paper's motivation and methodology.

2. If the paper was equity-related, it was further coded as whether it focused on equity-promoting opportunities ("opportunity-focused") or mitigating equity-related harms ("harm-focused"). The "opportunity-focused" category includes any paper that, broadly construed, applies an language-modeling-based system to increase social equity. The "harm-focused" papers focus on identifying or mitigating biases and other equity harms that may result from language models. For example, in (*3*), the authors co-design language modeling tools with activists to support the labor of tracking gender-based violence cases, and as such this paper is opportunity-focused. On the other hand, (*4*) benchmarks a language model's propensity to output harmful stereotypes, and is therefore harm-focused.

The three annotators exhibited high consistency, with Fleiss kappa interrater reliability statistics of 0.79 and 0.68 for the equity-related and opportunity-focused labels respectively, indicating substantial agreement (*5*). Out of the 100 manually annotated papers, 40 papers were identified as relevant to equity by at least one annotator. Of those, 6 (15%) papers were identified as opportunity-focused by at least one annotator, while the other 34 (85%) were not. When using majority vote (instead of at least one positive vote), the results were similar: 33 papers were majority annotated as equity-related, of which 5 (15%) were majority annotated as opportunity-focused vs. 28 (85%) as harm-focused. We conclude that in our sample of LLM papers, the overwhelming majority of equity-related papers are harm-focused as opposed to opportunity-focused.

[1] The only alteration to the keyword list used in (*2*) was that papers mentioning 'BERT' were not included because they produced many irrelevant results in the Semantic Scholar corpus.

# References

1. K. Lo, L. L. Wang, M. Neumann, R. Kinney, D. Weld, *Proceedings of the 58th Annual Meeting of the Association for Computational Linguistics*, D. Jurafsky, J. Chai, N. Schluter, J. Tetreault, eds. (Association for Computational Linguistics, Online, 2020), pp. 4969–4983.

2. R. Movva, *et al.*, Topics, Authors, and Networks in Large Language Model Research: Trends from a Survey of 17K arXiv Papers (2023).

3. H. Suresh, *et al.*, *Proceedings of the 2022 ACM Conference on Fairness, Accountability, and Transparency* (2022), pp. 667–678.

4. M. Nadeem, A. Bethke, S. Reddy, *Proceedings of the 59th Annual Meeting of the Association for Computational Linguistics and the 11th International Joint Conference on Natural Language Processing (Volume 1: Long Papers)*, C. Zong, F. Xia, W. Li, R. Navigli, eds. (Association for Computational Linguistics, Online, 2021), pp. 5356–5371.

5. M. L. McHugh, *Biochemia Medica* **22**, 276 (2012).